\documentclass[superscriptaddress,nofootinbib,preprint]{revtex4}
\usepackage{graphicx}
\usepackage{amssymb,amsmath,amsfonts}
\usepackage[hypertex]{hyperref}

\def\lozenge{\boxit{\hbox to 1.5pt{\vrule height 1pt width 0pt \hfill}}}
\def\ie{{\it i.e.}}
\def\eg{{\it e.g.}}
\def\etc{{\it etc}}

\def\mpl{\ifmmode M_{pl}\else $M_{pl}$\fi}
\def\mpl{\ifmmode \overline M_{Pl}\else $\bar M_{Pl}$\fi}
\def\to{\rightarrow}

\newcommand{\vev}[1]{\langle {#1} \rangle}
\newcommand{\ord}[1]{\mathcal{O}{(#1)}}

\newcommand{\gsim}{\gtrsim}

\newcommand{\met}{E_T\!\!\!\!\!\! /\,\,}

\newcommand{\dalam}{\raise-1mm\hbox{\large$\Box$}}

\newcommand{\beq}{\begin{equation}}
\newcommand{\eeq}{\end{equation}}

\begin{document}

\pagestyle{plain}

\hfill$\vcenter{
\hbox{\bf SLAC-PUB-14622}} $

\vskip 1cm
\title{Testing the OPERA Superluminal Neutrino Anomaly at the LHC}

\author{Hooman Davoudiasl}

\email{hooman@bnl.gov}
\affiliation{Department of Physics,
Brookhaven National Laboratory, Upton, NY 11973-5000, USA}
\author{{Thomas G. Rizzo}\footnote{Work supported in part
by the Department of Energy, Contract DE-AC02-76SF00515.}}
\email{rizzo@slac.stanford.edu}
\affiliation{SLAC National Laboratory, 2575 Sand Hill Rd.,
Menlo Park, CA  94025, USA}

\begin{abstract}

The OPERA collaboration has reported the observation of superluminal
muon neutrinos, whose speed $v_\nu$ exceeds that of light $c$, with
$(v_\nu - c)/c \simeq 2.5 \times 10^{-5}$.  In a recent work,
Cohen and Glashow (CG) have refuted this claim by noting that such neutrinos will lose energy, by pair-emission of
particles, at unacceptable rates.  Following the CG arguments,
we point out that pair-emissions consistent with the OPERA anomaly can lead to detectable signals for 
neutrinos originating from decays of
highly boosted top quarks at the LHC, allowing an independent test of the superluminal neutrino hypothesis.

\end{abstract}
\maketitle



The OPERA collaboration \cite{OPERA} has recently reported that the
muon neutrinos from the CERN CNGS beam travel the approximately
730~km distance to Gran Sasso Laboratory at a speed $v_\nu$ that
exceeds the speed of light $c$, with $(v_\nu - c)/c = [2.48 \pm
0.28\, ({\rm stat.}) \pm 0.30\, ({\rm sys.})] \times 10^{-5}$. This
result, if confirmed, would be one of the most significant
discoveries of recent times and would have important implications
for fundamental physics.  However, many observational considerations
point to severe tensions between the OPERA result and
well-established phenomena regarding neutrinos, such as flavor
oscillations and the Supernova (SN) 1987a explosion \cite{Coleman:1997xq,GSS}, 
among others \cite{pheno}.  It is thus
important to subject this ``anomaly" to as many tests as possible.

Cohen and Glashow \cite{CG} have very recently made the important
observation that if a neutrino can travel faster than light several
processes that are kinematically forbidden, within the current
Lorentz invariant framework, would become allowed.  In particular,
Ref.~\cite{CG} showed that bremsstrahlung processes in which
superluminal neutrinos emit particles in vacuo and lose energy would
lead to a severe depletion of the beam above about 12.5~GeV at Gran
Sasso, in stark conflict with the reported OPERA data \cite{OPERA}.
Assuming a flavor independent neutrino velocity $v_\nu$, with
negligible energy dependence, the leading process relevant for the
analysis in Ref.~\cite{CG} was $\nu_\mu \to \nu_\mu\, e^+ e^-$,
mediated by a $Z$ vector boson \cite{Bi:2011nd}.

Inspired by the Cohen-Glashow (CG) analysis, we propose that the
above vacuum bremsstrahlung processes could also lead to detectable
signals at the LHC, assuming that the energy independence of $v_\nu$
will persist up to energies of order 100~GeV.  This is a mild
assumption, given that the OPERA results show no significant energy
dependence up to about 50~GeV\footnote{In fact, consistency with the
SN 1897a data requires that $v_\nu$ steeply increase with energy,
above $\sim 10$~MeV.  However, we will assume that $v_\nu$ has
reached its asymptotic value near the GeV-scale and will not
increase further.}.  Our main observation is that decays of 
boosted top quarks at the LHC will contain neutrinos with energy
$E_\nu \gsim 100$~GeV and such neutrinos would undergo CG vacuum
bremsstrahlung processes, leading to the appearance of (multiple)
charged lepton pairs or jets, with significantly displaced vertices,
along the neutrino path.  

Although our proposal is based on the mechanism considered in Ref.~\cite{CG}, the 
physical parameters characterizing the LHC experiments can in principle yield 
independent and complementary tests of the OPERA  anomaly.  Obviously, 
the LHC would directly probe the vacuum bremsstrahlung 
hypothesis, by searching for the emitted particles.  In addition, given the high energy 
and the short travel distance (limited to the size of the detector) of the neutrinos, possible 
oscillation effects, say into sterile states with negligible coupling to the $Z$ boson, 
that could affect the efficiency of particle emission 
will not apply to our discussion.  While we do not assert that such a possibility 
is phenomenologically compelling, it serves as an example 
of how LHC searches could be valuable as 
independent tests of Lorentz invariance in the neutrino  
sector and, by extension, the OPERA results.  
We will next discuss the quantitative aspects of our proposal.

As pointed out in Ref.~\cite{CG}, the threshold for emission of a pair of 
particles $\bar \psi \psi$, is $E_0 = 2 m_\psi/\sqrt{v_\nu^2 - v_\psi^2}$, where $m_\psi$ is  
the mass of $\psi$ and $v_\psi$ is its maximum speed in vacuo; as in the CG framework, we 
assume conservation of energy and momentum \cite{AmelinoCamelia:2011bz}.  
For economy of assumptions, we  take $v_\psi = c$; we set $c=1$, henceforth.  Using the CG notation, we define 
$\delta \equiv v_\nu^2 -1$, and hence $E_0 = 2 m _\psi/\sqrt{\delta}$.  
We will set $\delta = 5.0 \times 10^{-5}$ hereafter.  For the emission of 
$e^+ e^-$, one then gets for the threshold energy $E_0(e) \simeq 145$~MeV.  Neglecting the mass of the electron, \ie, far 
above the energy threshold, the rate for emitting $e^+ e^-$ from a neutrino of energy $E_\nu$ was estimated in Ref.~\cite{CG} by 
\beq
\Gamma = k' \frac{G_F^2}{192 \pi^3} E_\nu^5 \delta^3\,,
\label{GamCG}
\eeq
where $k'=1/14$ and $G_F$ is the Fermi constant, assuming negligible 
vector coupling, $c_V^e$, for the electron to the $Z$ boson. 

\begin{figure}
\includegraphics[angle=90,width=0.64\textwidth]
{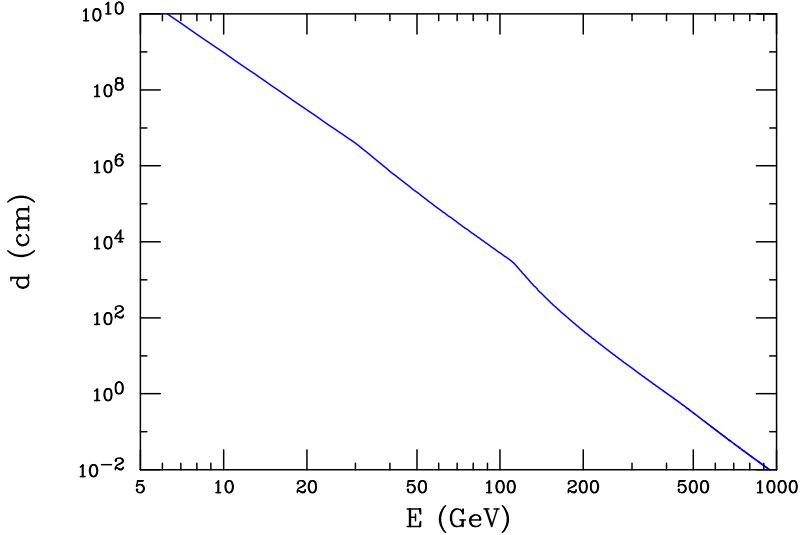}
\caption{Neutrino ``decay'' length, $d$, in units of centimeters as a function of neutrino energy in GeV.}
\label{fig:d}
\end{figure}

From the above Eq.~(\ref{GamCG}), we see that requiring the neutrino to have a 
``decay" length of $\sim 1$~m, so that the $e^+e^-$ pair emission could 
typically occur within an LHC detector, yields $E_\nu \sim 300$~GeV.  This suggests that 
a neutrino coming from the decay of a parent particle with $\sim 1$~TeV of energy would typically 
have one pair emission before leaving the detector volume.  Such a pair 
emission will lead to an energy loss of about $\sim 0.8 E_\nu$ \cite{CG}, 
dramatically decreasing the expected missing energy signal associated 
with a neutrino final state.  However, note that in such a case both the $e^+$ and $e^-$ would each 
carry more than $\sim 100$ GeV of energy making them easily visible in an LHC detector. 
We also note that for $E_\nu\sim 100-300$~GeV, as required here, other emission 
channels, in particular into $\mu^+ \mu^-$, $\bar u\, u$, and $\bar d \, d$ ({\it i.e.} jets) and possibly heavier quark pairs 
are also open.  A simple generalization of Eq.~(\ref{GamCG}), to include these additional channels, 
then yields (once their associated thresholds are passed)
\beq
\Gamma = k' \frac{G_F^2}{192 \pi^3} E_\nu^5 \delta^3 ~\sum_f ~N_c(f) ({c_V^f}^2+{c_A^f}^2)PS_f\,,
\label{Gam}
\eeq 
where the sum extends over kinematically accessible fermions, $PS_f$ represents the appropriate phase space suppression factor and $N_c(f)=1 (3)$ is a color factor for the fermion $f$.  
Here, $c_V^f$ and $c_A^f$ are the corresponding, appropriately normalized,  
vector and axial vector couplings of $f$ to the $Z$, respectively; 
\eg, $c_V^u=1-8x/3$, with $x=\sin^2\theta_W \simeq 0.23$.  
Note that fully including electron, muon and $u \bar u$ and $d\bar d$, \etc ~final states can increase the value of 
$\Gamma$ by more than an order of magnitude leading to a 
corresponding shortening of the decay length $d=\Gamma^{-1}$. In the calculations presented here 
we will assume that $PS_f \simeq (1-E_\nu^2/E_0^2(f))^{3/2}$, 
where $E_0(f)$ is the threshold energy for opening up ``decays'' into $\bar f f$. 
Other choices for $PS_f$ make very little difference in our numerical results except very close to thresholds.    
The result of this calculation over a wide range of neutrino energies is displayed in Fig.\ref{fig:d},  
where we have included all of the additional   
decay channels that open up when the various energy thresholds, $E_0(f)$, are crossed. For concreteness 
we have assumed that the relevant hadronic mass scales are those of $\rho$, $\phi$, $J/\psi$, and $\Upsilon$, 
corresponding to the emission of $u/d$, $s$, $c$, and $b$ quark pairs, respectively.  Thus, for example, ``decays'' to 
$\bar s s\; (\bar c c)$ have a neutrino threshold energy $E_0(f)$ of $m_{\phi\,(J/\psi)}/\sqrt \delta \simeq 144 \;(438)$ GeV.

\begin{figure}
\includegraphics[angle=90,width=0.64\textwidth]
{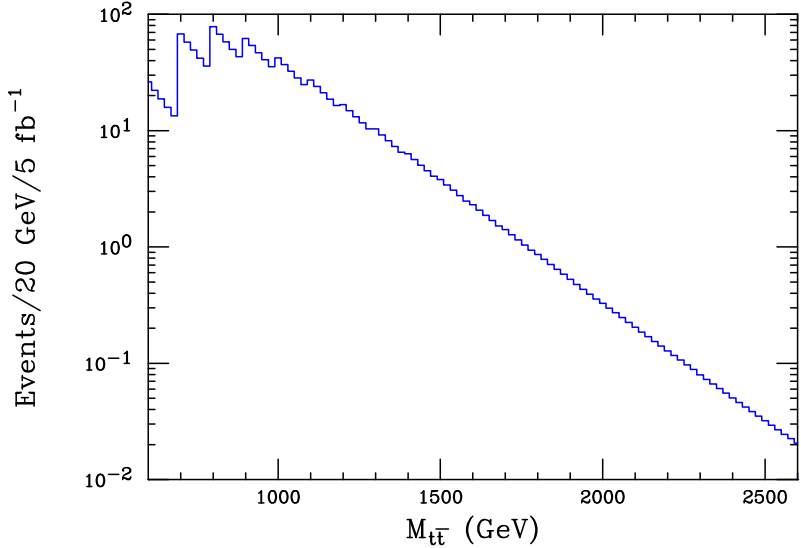}
\caption{$M_{t\bar t}$ distribution at the 7 TeV LHC for $\bar t t$ events, 
where one top decays leptonically, using the results of Ref.~\cite{joa}.}
\label{fig:mtt}
\end{figure}

Now let us address whether such a signal can be detected, in practice, at the LHC. 
Since the decay length for the neutrinos scales as $\sim 1/E^5$ and we are looking for a somewhat unusual process, 
it is advantageous to search for a copious production mechanism that leads to neutrinos of significantly high  
energy. One such likely possibility is $t \bar t$ production at large invariant masses, $M_{t\bar t}$,  
where one of the tops decays leptonically ({\it i.e.}, $t\to b \,W \to b \,\ell \,\nu_\ell; \, \ell=e,\mu$) 
and the resulting neutrino carries away $\sim 1/3$ of the parent $t$ (or $\bar t)$ energy. In such a 
case, neutrino energies in the interesting $\sim 100-300$ GeV energy range can easily result for $M_{t\bar t} \gsim 0.6-2$ TeV.  
We have plotted the $M_{t\bar t}$ distribution for these semileptonic top decay events per 5 fb$^{-1}$ in Fig.\ref{fig:mtt}, using the numerical  
results obtained from Ref.~\cite{joa}, that account for effects of top tagging efficiencies, leading to the 
modulated behavior for invariant masses below $\sim 1000$~GeV.  
In this figure, the next-to-leading order K-factors are included and a sum over both $e$ and $\mu$ final states has been performed.    
At the 7 TeV LHC with an integrated luminosity of $\sim 5$ fb$^{-1}$ 
this would typically yield a sample of $\sim 1400$ events with $M_{t\bar t}>600$ GeV.  Assuming an approximate equipartition of energy (which is verified by explicit calculations), so 
that $\vev{E_\nu}\sim  M_{t\bar t}/6$, the results shown in Fig.\ref{fig:mtt} can be used to 
estimate the corresponding distribution as a function of $E_\nu$.  Here we assume 
that both $\nu_e$ and $\nu_\mu$ are superluminal with equal speeds, as suggested by astrophysical and neutrino 
oscillation considerations \cite{GSS,CG,Coleman:1997xq}.  

In addition to the hadronic decay of the $t$ (or $\bar t$), the other side of the event would consist of the conventional $b$-jet 
plus a high-$p_T$ electron or muon. 
However, instead of the usual large $\sim 100-300$ GeV missing transverse energy 
$\met$ signal from the neutrino something 
different happens. After traveling $\sim 1$~m from the interaction point (IP)  
(the exact value depending on the neutrino energy as shown in 
Fig.\ref{fig:d}), 
the neutrino ``decays'' into a pair of leptons or into a hadronic jet plus a secondary neutrino, 
losing on average $\sim 80\%$ of its energy 
in the process \cite {CG}.   Since the initial neutrino is so highly boosted 
these decay products (which now also include the reduced 
$\met$ carried by the secondary neutrino) will approximately 
follow the original parent neutrino direction and leave tracks which point 
back to the IP but begin $\sim 1$m away at a displaced vertex \cite {searches}. 
Furthermore, if the parent neutrino energy is sufficiently 
high, there is a small chance that this process may be observed 
{\it twice} in the detector when the secondary neutrino again ``decays'', 
producing an extraordinary signature.  However, this signature is most likely not  
relevant for the early run at the 7 TeV LHC, as we will discuss below.  

Taking $d \simeq 10$~m as the maximum 
allowed decay length by typical dimensions of an LHC detector, Fig.\ref{fig:d} suggests that the 
secondary neutrino, after the first emission depletes the initial energy by about 80\% \cite{CG}, 
should have $E_\nu \gsim 100$~GeV.  This  
would require an initial neutrino energy larger than about 500~GeV.  Under the 
assumption of energy equipartition for typical top decays, we then would require  $M_{{\bar t} t} \gsim 3000$~GeV.  
Hence, as indicated by the distribution in Fig.~\ref{fig:mtt}, the 7 TeV run of the LHC would require 
a data sample much larger than $\ord{10}$ fb$^{-1}$ in order to search for the double-emission signal.  
Seeing this effect would likely require the 14 TeV LHC at full design luminosity. 

Finally, we would like to add that 
when both the $t$ and $\bar t$ decay leptonically, the neutrino decay processes will typically  
take place on both sides of the $t \bar t$ event.  However, in this case, due to the smaller branching fraction 
the statistics will be reduced by a factor of about 6.

In conclusion, if the OPERA anomaly \cite{OPERA} is due to superluminal neutrinos, the results of Ref.~\cite{CG}
imply that distinct signals should appear in the decays of highly boosted top quarks (unless 
the superluminal behavior becomes less prominent above the OPERA beam energies; 
this would seemingly require such behavior to be only limited to neutrino energies of roughly 1-100~GeV).  Our 
estimates suggest  that the currently available center of mass energy (7~TeV) and integrated luminosity 
($\sim 5$ fb$^{-1}$) should be sufficient to test the OPERA anomaly at the LHC.

\acknowledgments

The work of H.D. is supported by the DOE grant DE-AC02-98CH10886. T.G.R. would like to thank J. Hewett for access to some of the detailed 
results of Ref.\cite{joa}.

\end{document}